\documentclass[a4paper,11pt]{article}
\usepackage{pos}
\usepackage{multirow}
\usepackage{float}
\usepackage{amsfonts}
\usepackage{dsfont}
\usepackage{graphicx}
\usepackage{scrhack}
\usepackage{float}
\usepackage[section, below]{placeins}

\usepackage{subcaption}
\usepackage{mdframed}

\newcommand{\etal}{\textit{et al.}}

\title{$T_{cc}$  pole trajectory}

\author*[a]{Protick Mohanta}
\author[b,c]{Srijit Paul}
\author[d,e]{Subhasish Basak}

\affiliation[a]{The Institute of Mathematical Sciences, Chennai 600113, India}

\affiliation[b]{Maryland Center for Theoretical Physics, University of Maryland, College Park, MD 20742 USA }
\affiliation[c]{Department of Physics, University of Cyprus, Aglantzia 2109, Nicosia, Cyprus}

\affiliation[d]{School of Physical Sciences, National Institute of Science Education and Research, An OCC of Homi Bhabha National Institute, Jatni 752050, India}

\affiliation[e]{Homi Bhabha National Institute, Training School Complex,
Anushakti Nagar, Mumbai  400094, India}

\emailAdd{protickm@imsc.res.in}
\emailAdd{spaul137@umd.edu}
\emailAdd{sbasak@niser.ac.in}

\abstract{ We investigate the spectrum of doubly charmed
tetraquark $T_{cc}$ with quantum number $I(J^P) = 0(1^+)$
using MILC's $N_f = 2+1+1$ HISQ gauge ensembles at two lattice
spacings. We have included diquark-antidiquark operator together
with molecular and scattering operators in our analysis and
varied both the heavy and light quark masses. We employ the
anisotropic Clover action for heavy quarks, and $O(a)$-improved
Wilson--Clover action for the light (up/down) quarks. In order
to handle the non-analyticity near the Left Hand Cut we use
modified L\"uscher's method when we are close to it.}

\FullConference{The 42nd International Symposium on Lattice Field Theory (LATTICE 2025)\\
2-8 November 2025\\
Tata Institute of Fundamental Research, Mumbai, India\\}

\begin{document}
\maketitle

\section{Introduction}
The discovery of doubly charmed tetraquark $T_{cc}$~\cite{Aaij}
has led to extensive investigations, both lattice and non-lattice,
of doubly charm ($T_{cc}$), doubly bottom ($T_{bb}$) and other
possible heavy tetraquark candidates like $T_{bc}$, $bb\bar{u}
\bar{s}$ etc. Though lattice QCD mostly indicates the existence
of deeply bound doubly bottom tetraquark $T_{bb}(bb\bar{u}\bar{d})$
but the theoretical prediction is less certain about the existence
of doubly charm tetraquark $T_{cc}(cc\bar{u}\bar{d})$ ~\cite{Chen}.
The pole of $T_{cc}$ lies a mere 0.36 MeV below the
$D^0 D^{\ast\,+}$
threshold, whereas the binding energy of $T_{bb}$ is $\sim O(100)$
MeV~\cite{francis1,francis2,junnarkar,luka,Mohanta1}. As a first
step to address this, we have varied the heavy quark mass and taken
three additional mass points between charm and bottom mass. The
existence of Left Hand Cut (LHC) complicates the situation as one
approaches the physical pion mass limit~\cite{Du_PRL}. Therefore
while varying the light quark mass, we used L\"uscher method
when we are away from LHC and modified L\"uscher method~\cite{mod_luscher}
when closer to it. In this proceeding we report our preliminary result
and the status of our ongoing $T_{cc}$ pole analysis.

\section{Operator Basis}
Many early studies such as~\cite{Padmanath1}~ have not included
diquark-antidiquark operator in their analysis. Cheung \etal
~\cite{Cheung}~ found diquark-antidiquark operator not to have
significant effects on finite volume spectra and Cheng \etal
\cite{Cheng}~ showed such operator results in an unstable $T_{cc}$.
Most of the past lattice {$T_{cc}$} investigations thus did not
consider diquark-antidiquark operator in their simulations. However,
studies based on heavy quark symmetries \cite{Eichten,Mehen1}~
showed the usefulness of diquark operators in doubly heavy tetraquark
states. Another striking point is the mass differences in $J^P=(1/2)^+$
singly heavy baryons $(\Lambda_b, \Sigma_b)$, $(\Lambda_c, \Sigma_c)
\sim 191-167$ MeV. The operators 
\begin{equation}
\left(\mathcal{O}^{h l_1 l_2}_k \right)_\alpha = \epsilon_{abc}
\left[{l_1^a}^T C\gamma_k l_2^b \right]\, Q^c_\alpha \;\;\;\text{and}
\;\;\; \left(\mathcal{O}^{h l_1 l_2}_5 \right)_\alpha = \epsilon_{abc}
\left[{l_1^a}^T C \gamma_5 l_2^b \right]\, Q^c_\alpha \label{lam_sig_op}
\end{equation}
gives a mass difference of $\Sigma_b - \Lambda_b \sim 30$ MeV at a
pion mass of 490 MeV. When we reduce the pion mass, the above two
operators in eqn. (\ref{lam_sig_op}) produce significantly enhanced
mass splitting \cite{Bowler}. And since the diquark-antidiquark
operator is related to the $\Lambda_Q$ operator by heavy quark-diquark
symmetry, we expect it to have important contribution to $T_{cc}$. We,
therefore, use the following operators in our analysis
\begin{eqnarray}
\mathcal{D}(x) &=& \left[ c(x)^{a\,T} C\gamma_k \,c(x)^b \right]\,\left[
\bar{u}(x)^a \,C\gamma_5 \,\bar{d}(x)^{b\,T} \right] \nonumber \\
\mathcal{M}_1(x) &=& \left[ \bar{d}(x)^a \,\gamma_k \,c(x)^a \right] \,
\left[\bar{u}(x)^b \,\gamma_5 \,c(x) ^b \right] \nonumber \\
\mathcal{M}_2(x) &=& \epsilon_{kij}\left[ \bar{d}(x)^a \,\gamma_i \,c(x)^a
\right] \, \left[\bar{u}(x)^b \,\gamma_j \,c(x) ^b \right] \nonumber \\
\mathcal{S}(t;\mathbf{p}_1, \mathbf{p}_2) &=& \sum_{\mathbf{x}}
\left[ \bar{d}(x)^a\,\gamma_k \,c(x)^a \right] \,e^{i\mathbf{p}_1\cdot
\mathbf{x}} \,\times\,\sum_{\mathbf{y}} \left[\bar{u}(y)^b \,\gamma_5 \,
c(y)^b \right] \,e^{i\mathbf{p}_2 \cdot \mathbf{y}} \label{operators}
\end{eqnarray}
We perform our analysis in the centre-of-mass frame, hence
the $D^\ast$ and $D$ mesons in the scattering operator $\mathcal{S}$
are given back-to-back momentum $\mathbf{p}_1 + \mathbf{p}_2 =0$.
We have generated data up to $\mathbf{p}^2=1$ and performed GEVP
analysis of the $5 \times 5$ correlator matrix. Here we present three
diagonal elements of GEVP matrix, $\mathcal{C}_{\mathcal{DD}}(t)$,
$\mathcal{C}_{\mathcal{M}_i\mathcal{M}_i}(t)$ and $\mathcal{C}_{
\mathcal{SS}}(t)$,
\begin{eqnarray}
\mathcal{C}_{\mathcal{DD}}(t) &=& \sum_{\vec{x}} \left\langle
\mathcal{D}(x) \,\mathcal{D}(0)^\dagger \right\rangle \nonumber \\
 &=& \;\;\;\;\sum_{\vec{x}} \textmd{Tr} \Big[ \Big\{
 \mathcal{G}_c(t,\vec{x};0) \Big\}^{ad\,T} \; \Big( \gamma_k \gamma_4
 \gamma_2 \mathcal{G}_c (t,\vec{x};0) \gamma_4 \gamma_2 \gamma_k
 \Big)^{bc}\, \Big] \nonumber\\
 && \hspace{0.4in} \times \;\textmd{Tr} \Big[ \Big\{\gamma_4 \gamma_2
 \mathcal{G}_u(t,\vec{x};0)^\dagger  \gamma_4 \gamma_2 \Big\}^{da}
\left(\gamma_5 \mathcal{G}_d(t,\vec{x};0)^\dagger \gamma_5
\right)^{cb\,T} \Big] \nonumber \\
 && - \sum_{\vec{x}} \textmd{Tr} \Big[ \Big\{ \mathcal{G}_c(t,\vec{x};0)
\gamma_4 \gamma_2\gamma_k \Big\}^{ac} \; \Big(\gamma_k \gamma_4 \gamma_2
\mathcal{G}_c(t,\vec{x};0) \Big)^{bd\,T} \Big] \nonumber\\
&& \hspace{0.4in} \times \;\textmd{Tr} \Big[ \left\{\gamma_4
\gamma_2  \mathcal{G}_u(t,\vec{x};0)^\dagger  \gamma_4 \gamma_2
\right\}^{da}\; \left(\gamma_5 \mathcal{G}_d(t,\vec{x};0)^\dagger
\gamma_5 \right)^{cb\,T}
\Big]  \label{corr_dd}
\end{eqnarray}
\begin{eqnarray}
\mathcal{C}_{\mathcal{M}_1\mathcal{M}_1}(t) &=& \sum_{\vec{x}}
\left\langle
\mathcal{M}_1(x) \,\mathcal{M}_1(0)^\dagger \right\rangle \nonumber \\
 &=& \;\;\;\;\sum_{\vec{x}} \textmd{Tr} \Big[ \Big\{\gamma_5
\mathcal{G}_d(t,\vec{x};0)^\dagger \gamma_5\Big\} \,\Big(\gamma_k
\mathcal{G}_c(t,\vec{x};0)
\gamma_k \big) \Big] \, \textmd{Tr} \Big[\Big\{ \gamma_5
\mathcal{G}_u(t,\vec{x};0)^\dagger \gamma_5 \Big\} \,\Big(\gamma_5
\mathcal{G}_c(t,\vec{x};0)
\gamma_5 \Big)\Big] \nonumber \\
 && - \sum_{\vec{x}} \textmd{Tr} \Big[ \Big\{\gamma_5
 \mathcal{G}_d(t,\vec{x};0)^\dagger
\gamma_5 \Big\}\,\Big( \gamma_k \mathcal{G}_c(t,\vec{x};0) \gamma_5
\Big) \,\Big\{ \gamma_5 \mathcal{G}_u(t,\vec{x};0)^\dagger
\gamma_5 \Big\} \,\Big( \gamma_5
\mathcal{G}_c(t,\vec{x};0) \gamma_k \Big) \Big] \label{corr_m1m1} 
\end{eqnarray}
\begin{eqnarray}
 \mathcal{C}_{\mathcal{SS}}(t;\mathbf{p}_1,\mathbf{p}_2,\mathbf{p}_4)
 &=& \left\langle \mathcal{S} (t;\mathbf{p}_1,\mathbf{p}_2) \,
 \mathcal{S}(0;\mathbf{p}_3,\mathbf{p}_4)^\dagger \right\rangle \nonumber\\
&=& \sum_{\mathbf{x},\mathbf{y},\mathbf{z}} e^{i(\mathbf{p}_1\cdot\mathbf{x}
+ \mathbf{p}_2 \cdot \mathbf{y} - \mathbf{p}_4 \cdot \mathbf{z})} \;
\textmd{Tr} \Big[ \Big\{\gamma_5 \mathcal{G}_d(t,\mathbf{x};0)^\dagger
\gamma_5 \Big\} \, \Big( \gamma_k \mathcal{G}_c(t,\mathbf{x};0)
\gamma_k \Big) \Big] \nonumber \\
 && \hspace{0.9in} \times \;\textmd{Tr} \Big[
 \mathcal{G}_c(t,\mathbf{y};0,\mathbf{z})\Big\{ \gamma_5
 \mathcal{G}_u(0,\mathbf{z};t,\mathbf{y})\gamma_5  \Big\}\Big] \nonumber \\
 && - \sum_{\mathbf{x},\mathbf{y},\mathbf{z}} e^{i(\mathbf{p}_1  \cdot
 \mathbf{x} + \mathbf{p}_2\cdot\mathbf{y}- \mathbf{p}_4\cdot\mathbf{z})} \;
 \textmd{Tr} \Big[\Big\{\gamma_k\gamma_5 \mathcal{G}_d(t,\mathbf{x};0)
 ^\dagger \gamma_5 \gamma_k \Big\} \Big(
 \mathcal{G}_c(t,\mathbf{x};0,\mathbf{z})\nonumber \\
 && \hspace{1.45in} \gamma_5 \mathcal{G}_u(0,\mathbf{z};t,\mathbf{y})
  \,  \gamma_5 \mathcal{G}_c(t,\mathbf{y};0)\Big) \Big] \label{corr_ss}
\end{eqnarray}
Except for the correlator $\mathcal{C}_{\mathcal{SS}}(t;\mathbf{p}_1,
\mathbf{p}_2,\mathbf{p}_4) = \left\langle \mathcal{S} (t;\mathbf{p}_1,
\mathbf{p}_2) \,\mathcal{S}(0;\mathbf{p}_3,\mathbf{p}_4)^\dagger
\right\rangle $, all other require computation of only point-all propagators.
For the $\mathcal{C}_{\mathcal{SS}}$ we made use of the
one end trick as suggested in \cite{Dina}. In this case, we make use of
translational invariance to remove the phase related to
$\mathbf{p}_3$. We implemented complex $\mathbb{Z}(2)\times \mathbb{Z}
(2)$ random numbers at time slice $t=0$ for inversion of the Dirac
operators corresponding to charm and up/down quarks. For instance, the
first term of the eqn. (\ref{corr_ss}) can be written as
\begin{eqnarray}
&& {1\over N}\sum_n\sum_{\mathbf{x}}e^{i(\mathbf{p}_1  \cdot
\mathbf{x})} \Big\{\gamma_k\gamma_5 \mathcal{G}_d(t,\mathbf{x};0)^\dagger
\gamma_5 \gamma_k\Big\}^{c_1c_2}_{s_1s_2} \, \Big(\mathcal{G}_c(t,
\mathbf{x};0) \Big)^{c_2c_1}_{s_2s_1} \nonumber \\
&& \;\;\;\times \;\; \sum_{\mathbf{y}}e^{i(\mathbf{p}_2  \cdot \mathbf{y})}
\Big(\phi_c^n(\mathbf{y},t) \Big)^{c_3}_{s_3}\Big(\phi_u^n(\mathbf{y},
t)^\dagger\Big)^{c_3}_{s_3} \label{frst_term1}, \\
\text{where,} && \nonumber \\
&& \; \Big(D_c(r,x)\Big)^{c_1c_2}_{s_1s_2}\Big(\phi_c^n(x)\Big)^{c_2}_{s_2}
 =\delta_{r_0,0}\Big(\Xi(\mathbf{r})[n]\Big)^{c_1}_{s_1} \nonumber \\
&& \; \Big(D_u(r,x)\Big)^{c_1c_2}_{s_1s_2}\Big(\phi_u^n(x)\Big)^{c_2}_{s_2}
 =\delta_{r_0,0}\Big(\Xi(\mathbf{r})[n]\Big)^{c_1}_{s_1}e^{i(\mathbf{p}_4
\cdot \mathbf{r})}\label{one_end_1}
\end{eqnarray}

\section{Quark action and tuning}

\subsection{Clover action for light up/down quark}

To simulate light up/down quarks, we employ the $\mathcal{O}(a)$
improved Wilson--Clover fermion action, which includes a clover leaf
term that systematically removes leading discretization error at
$\mathcal{O}(a)$. The corresponding fermion action is expressed as
\begin{eqnarray}
S_{\textmd{clover}} &=& \sum_n \bar{\psi}(n)\psi(n) - \kappa_s\Big[
\sum_{n,\mu}\bar{\psi}(n)(1-\gamma_\mu)U_\mu(n)\psi(n+\hat{\mu})\nonumber\\
&+& \sum_{n,\mu}\bar{\psi}(n)(1+\gamma_\mu)U^\dagger_\mu(n-\hat{\mu})
\psi(n-\hat{\mu})\Big] -\kappa_s c_{\textmd{SW}}\sum_{n,\mu<\nu}
\bar{\psi}(n)\sigma_{\mu\nu}F_{\mu\nu}(n)\psi(n)
 \label{clover}
\end{eqnarray}
where $\kappa_s =\dfrac {1} {2(m + 4)}$ and $c_{\text{SW}}$ is the
clover coefficient. We vary the pion mass from $m_{\eta_s}=
688.5$ MeV \cite{Dowdall:2013rya} down to 400 MeV, where $m_{\eta_s}$
denotes the mass of the fictitious $\eta_s$ meson \cite{Mohanta:2019mxo}.
The gauge links are HYP smeared and $c_{\text{SW}}$ is obtained from 
\cite{FermilabLattice:2011njy}
\begin{equation}
 c_{\text{SW}} = \frac{1}{u_0^3},  \label{csw}
\end{equation}
where $u_0$ is the tadpole improvement factor defined by the fourth
root of the average plaquette. We tabulate the pion masses used
in our analysis in Table \ref{tab:lightk}.

\subsection{Anisotropic clover action for charm quark} \label{anisoclv}

In this work, we use the anisotropic clover-improved Wilson action
\textit{i.e.} Relativistic Heavy Quark (RHQ) action
\cite{rhq1,rhq2} for the charm
quark. The RHQ action modifies the standard Wilson-Clover action
by introducing an anisotropy factor $\nu$. This enhances temporal
resolution which is crucial for heavy quark systems on lattices
with shorter time direction 
sizes. The RHQ action $S_{\textmd{RHQ}}$ is given by
\begin{eqnarray}
{S_{\textmd{RHQ}} \over m_0+1+3\nu} &=& \sum_n \bar{\psi}(n)\psi(n) -
\kappa\Big[\sum_{n}\bar{\psi}(n)(1-\gamma_0)U_0(n)\psi(n+\hat{0})
\nonumber\\
&+& \sum_{n}\bar{\psi}(n)(1+\gamma_0)U^\dagger_0(n-\hat{0})
\psi(n-\hat{0}) + \nu\sum_{n,i}\bar{\psi}(n)(1-\gamma_i)U_i(n)
\psi(n+\hat{i})\nonumber\\
&+& \nu\sum_{n,i}\bar{\psi}(n)(1+\gamma_i)U^\dagger_i(n-\hat{i})
\psi(n-\hat{i})\Big] -\kappa c_P\sum_{n,\mu<\nu}
\bar{\psi}(n)\sigma_{\mu\nu}F_{\mu\nu}(n)\psi(n), \nonumber \\
 && \label{rhq}
\end{eqnarray}
where  $\kappa =  1/[2(m_0 +1+3\nu)]$, $c_P$ is the clover
coefficient, $\sigma_{\mu\nu} = \frac{i}{2} [\gamma_\mu, \gamma_\nu]$
is the antisymmetric tensor and $F_{\mu\nu}(n)$ represents the
discretized gluon field strength. First we tuned the strange propagator
using Wilson-Clover action (\ref{clover}) and subsequently tuned
the RHQ action parameters $\{ m_0, c_P, \nu \}$. The $m_0$ is tuned
by matching spin-average $aM_{D_s}$ and $aM_{D_s^\ast}$ obtained
from lattice with the corresponding experimental value
$\overline{M}_{D_s}$
\begin{equation}
\overline{M}_{D_s} = \frac{1}{4} M_{D_s} + \frac{3}{4} M_{D_s^*}
\;\approx \; 2.076 \;\text{GeV} \label{spinavDs}
\end{equation}
The clover coefficient $c_P$ is adjusted so that the hyperfine
splitting matches its PDG value,
\begin{equation}
\Delta M_{D_s} = M_{D_s^*} - M_{D_s} \; \approx \; 143.8 \;
\text{MeV}. \label{deltaDs}
\end{equation}
The anisotropy coefficient $\nu$ is tuned to satisfy relativistic
dispersion relation
\begin{equation}
E^2(\vec{p}) = M^2 + \vec{p}^{\,2}, \label{reldisper}
\end{equation}
ensuring the speed of light is normalized on the lattice to $c^2=1$.
We used $|\vec{p}|={2\pi \over L}$ in tuning the RHQ action parameters.

\section{Simulation Details}
The simulations have been performed using two ensembles of MILC
$N_f=2+1+1$ HISQ lattices~\cite{MILC:2010pul}, the details of which
are given in Table \ref{tab:ensembles}. The parameter tuning and
production run on $24^3 \times 64$ lattices are ongoing, here we
present a status update of the project. All measurements are carried
out using point source.
\begin{table}[h]
\centering
\begin{tabular}{|c|c|c|c|c|c|c|}
\hline
$\beta = {10} /{g^2}$ & $m_l$ & $m_s$ & $m_c$ &
$L^3 \times T$ & $a$ (fm) & $N_{cfg}$ \\ \hline
5.80 & 0.013 & 0.065 & 0.838 & $16^3 \times 48$ & 0.15 & 700 \\ \hline
6.00 & 0.0102 & 0.0509 & 0.635 & $24^3 \times 64$ & 0.12 & 700 \\ \hline
\end{tabular}
\caption{MILC configurations used in this work and $N_{cfg}$ is the
number of configurations used.}
\label{tab:ensembles}
\end{table}

\subsection{Heavy-light meson mass}

\begin{center}
\begin{figure}[H]
\centerline{\includegraphics[scale=0.54]{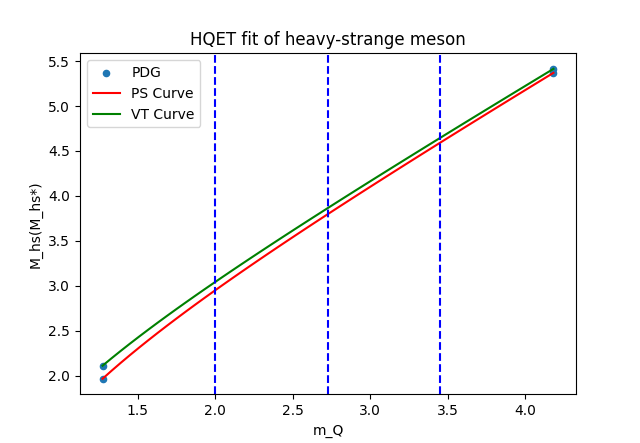}}
\caption{Heavy-light meson points}
\label{fig:hq_mass}
\end{figure}
\end{center}

With the varying of heavy quark mass, we need both the heavy-strange
pseudoscalar and vector meson masses for the tuning of RHQ action
parameters. HQET provides the relevant expression for the masses
of heavy-light mesons \cite{Manohar1}
\begin{equation}
M_{hl} = m_h + \bar{\Lambda} -\frac{\lambda_1}{2m_h} -\frac{3
\lambda_2}{2m_h} \;\;\; \text{and} \;\;\; M_{hl}^\ast = m_h +
\bar{\Lambda} -\frac{\lambda_1}{2m_h} + \frac{\lambda_2}{2m_h}
\label{man_hl}
\end{equation}
In figure \ref{fig:hq_mass} we plot our heavy-strange pseudoscalar
$M_{hl}$ and vector meson $M_{hl}^\ast$ mass points, starting from
$D_s,\; D_s^\ast$ (the two leftmost points) all the way to $B_s,\;
B_s^\ast$ (the two rightmost points).

\subsection{Light quark tuning}

For the strange quark, the $\kappa_s$ has been tuned separately on
each ensemble to arrive at ChiPT $\bar{s} \gamma_5 s$ state $\eta_s
= 688.5$ MeV mass as discussed before. The rest of the $\kappa$
values are chosen to get the pion masses approximately equidistant
between successive points. To avoid uncontrolled fluctuations, we
limited ourselves to $m_\pi \ge 400$ MeV. We wish to point out that
without the HYP-smearing the GEVP matrix looses definite positivity
at around time slice $t=4$.
\begin{table}[h]
\begin{tabular}{|c|c|ccccccc|} \hline
 \multirow{2}{*}{$16^3 \times 48$} & $\kappa$ & 0.12566 & 0.1260 &
0.1263 & 0.1266 & 0.1269 & 0.1272 & 0.1275\\
 & $m_\pi$  & 688.8(3) & 644.7(3) & 603.3(3) & 559.3(3) & 511.7(3) &
459.1(3) & 400.4(5) \\ \hline
\multirow{2}{*}{$24^3 \times 64$} & $\kappa$ & 0.1256 & 0.125845 &
0.126062 & 0.12628 & 0.126498 & 0.126718 & 0.126937\\
 & $m_\pi$  & 688.7(2) & 644.7(2) & 603.2(2) & 559.2(2) & 511.7(2) &
459.2(2) & 400.7(2) \\ \hline
\end{tabular}
\caption{The light $\kappa$ used in our analysis. The $m_\pi$ are
in MeV.} \label{tab:lightk}
\end{table}

\subsection{RHQ parameters tuning}
\begin{figure}[H]
\vspace{-0.2in}
    \begin{subfigure}[t]{0.55\textwidth}
\includegraphics[width=\textwidth, height=0.249\textheight]
{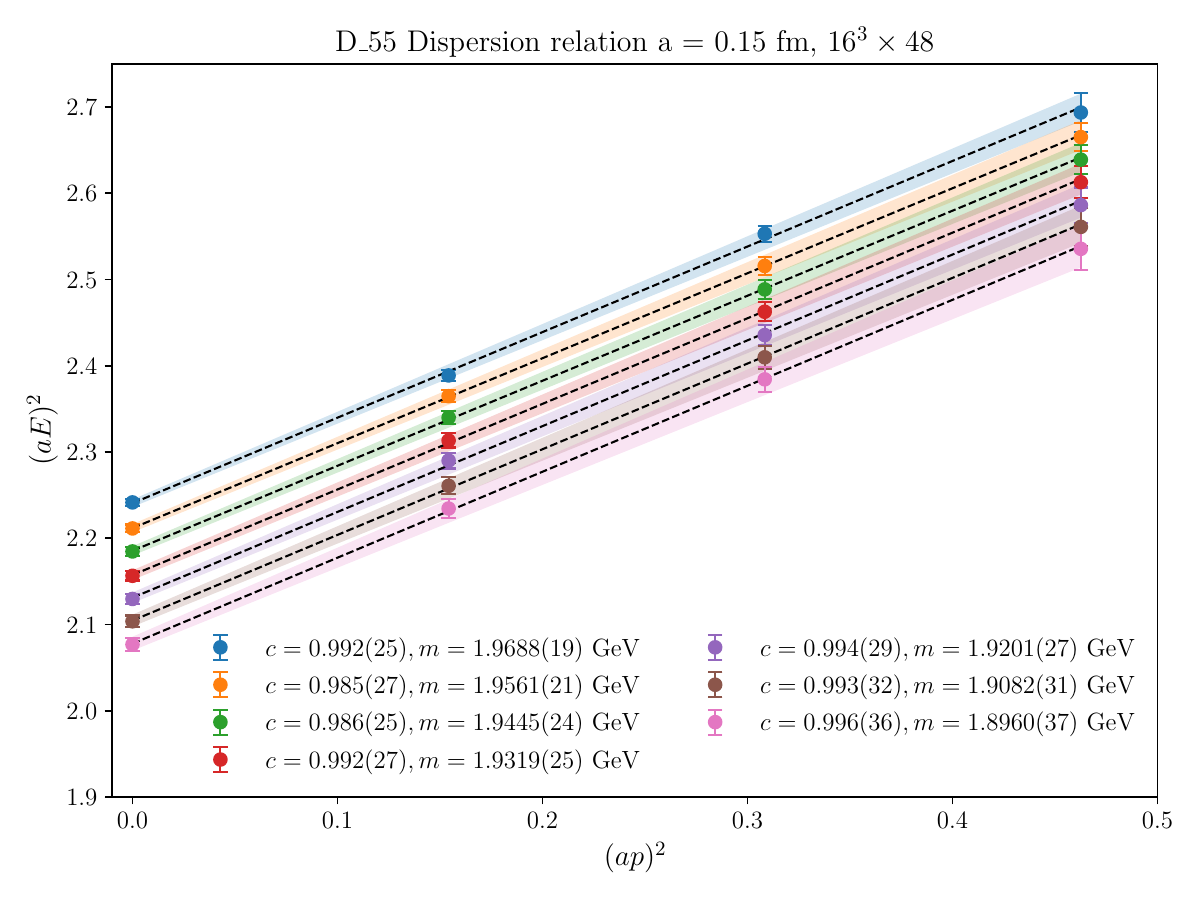}
    \end{subfigure}%
    \begin{subfigure}[t]{0.55\textwidth}
\includegraphics[width=\textwidth, height=0.249\textheight]
{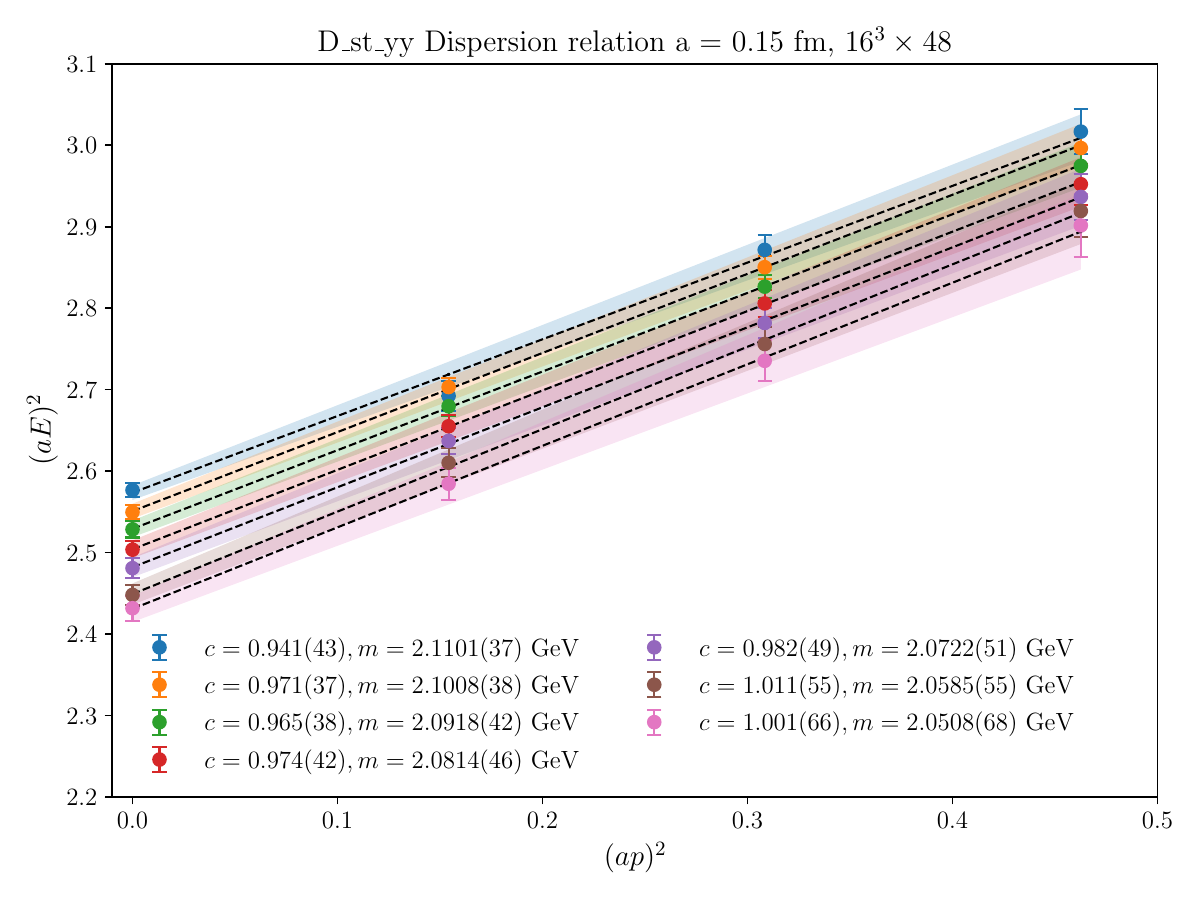}
    \end{subfigure}
    \caption{Dispersion relation of $D_s$ and $D_s^\ast$ meson obtained
    on $16^3 \times 48$ ensemble}
    \label{fig:dispersion1}
\end{figure}

The step to tune charm quark involves determining $m_0$ along with
the other two RHQ action parameters $\{c_P,\nu\}$ appearing in
(\ref{rhq}). As discussed in subsection \ref{anisoclv}, tuning
involve matching the spin-averaged mass, the hyperfine splitting
and the velocity of light $c^2=1$ for the $m_Q$ points obtained
from eqn. (\ref{man_hl}) and the plot Fig. \ref{fig:hq_mass}.
The tuned RHQ parameters for $16 \times 48$ are given in Table
\ref{tab:RHQ_param}. In figure \ref{fig:dispersion1} we present
the dispersion relation of $D$ and $D^\ast$ for $\kappa = 0.12566$.
\begin{table}[h]
\begin{center}
\begin{tabular}{|ccccccc|} \hline 
$m_h$  & $m_0$ & $\nu$ & $C_p$ & $\overline{M}_{hs}$ &
$\Delta M_{hs}$ & $c^{\text{latt}}$ \\
\hline
1.273  & 1.462 & 1.351 & 1.976 & 2076.8(9) & 144.6(9) & 0.992(25) \\
2.00  & 4.009 & 2.036 & 2.710 & 3020.5 (9) & 93.5 (9) & 1.021 (23) \\
2.73  & 8.499 & 2.951 & 3.922 & 3853.0 (11) & 69.8 (11) & 1.004 (27) \\
3.45  & 16.244 & 4.461 & 5.953 & 4631.4 (13) & 57.5 (13) & 1.001 (29)\\
4.183  & 30.061 & 7.041 & 9.014 & 5403.0(14) & 48.3(14) & 1.006 (30) \\
\hline
\end{tabular}
\caption{RHQ action parameter tuned on $16^3 \times 48$ ensemble. The
$\overline{M}_{hs}$ and $\Delta M_{hs}$ are in MeV.}
\label{tab:RHQ_param}
\end{center}
\end{table}

\section{Results}

\begin{figure}[!ht]
    \begin{subfigure}[t]{.32\textwidth}
\includegraphics[scale=0.30]{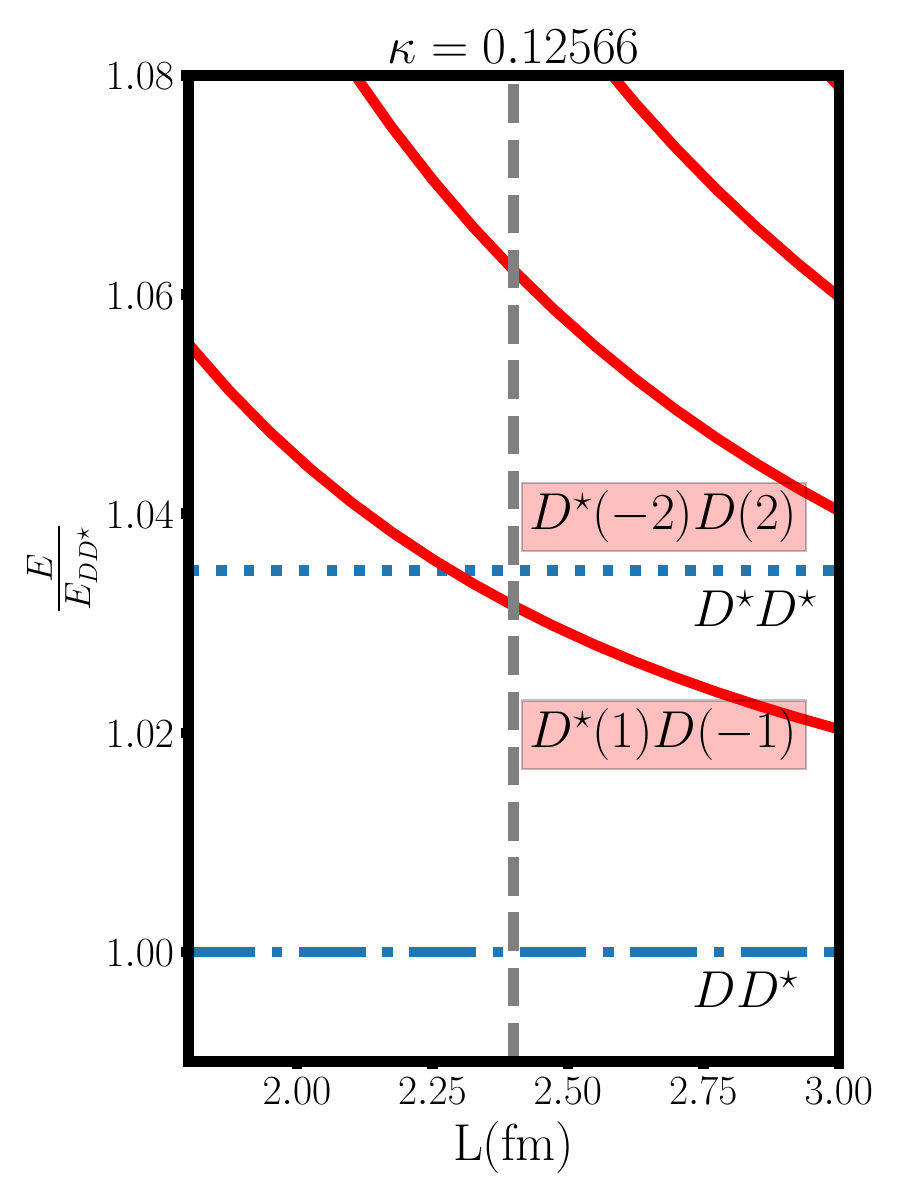}
\caption{$\kappa$=0.12566}
    \end{subfigure}%
    \begin{subfigure}[t]{.32\textwidth}
\includegraphics[scale=0.30]{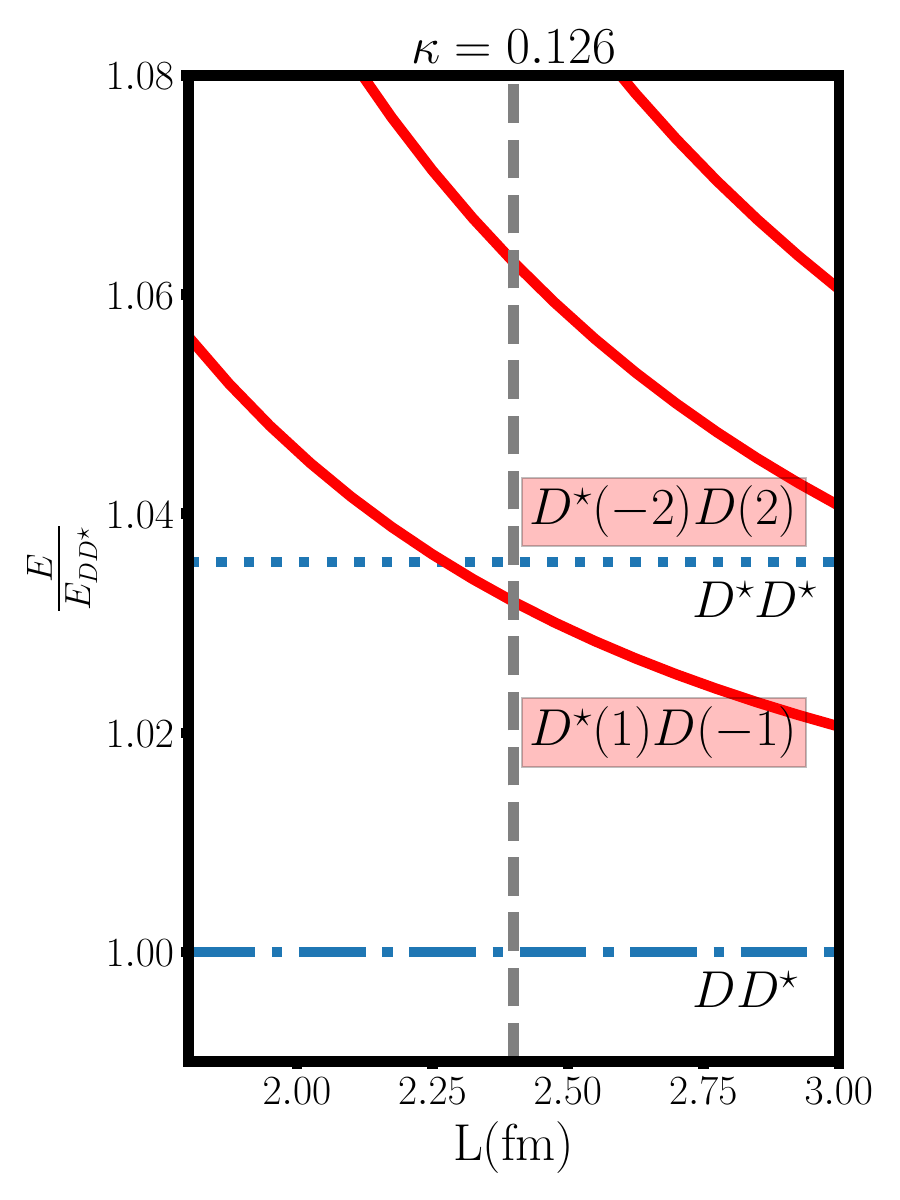}
\caption{$\kappa$=0.1260}
    \end{subfigure}%
    \begin{subfigure}[t]{.32\textwidth}
\includegraphics[scale=0.30]{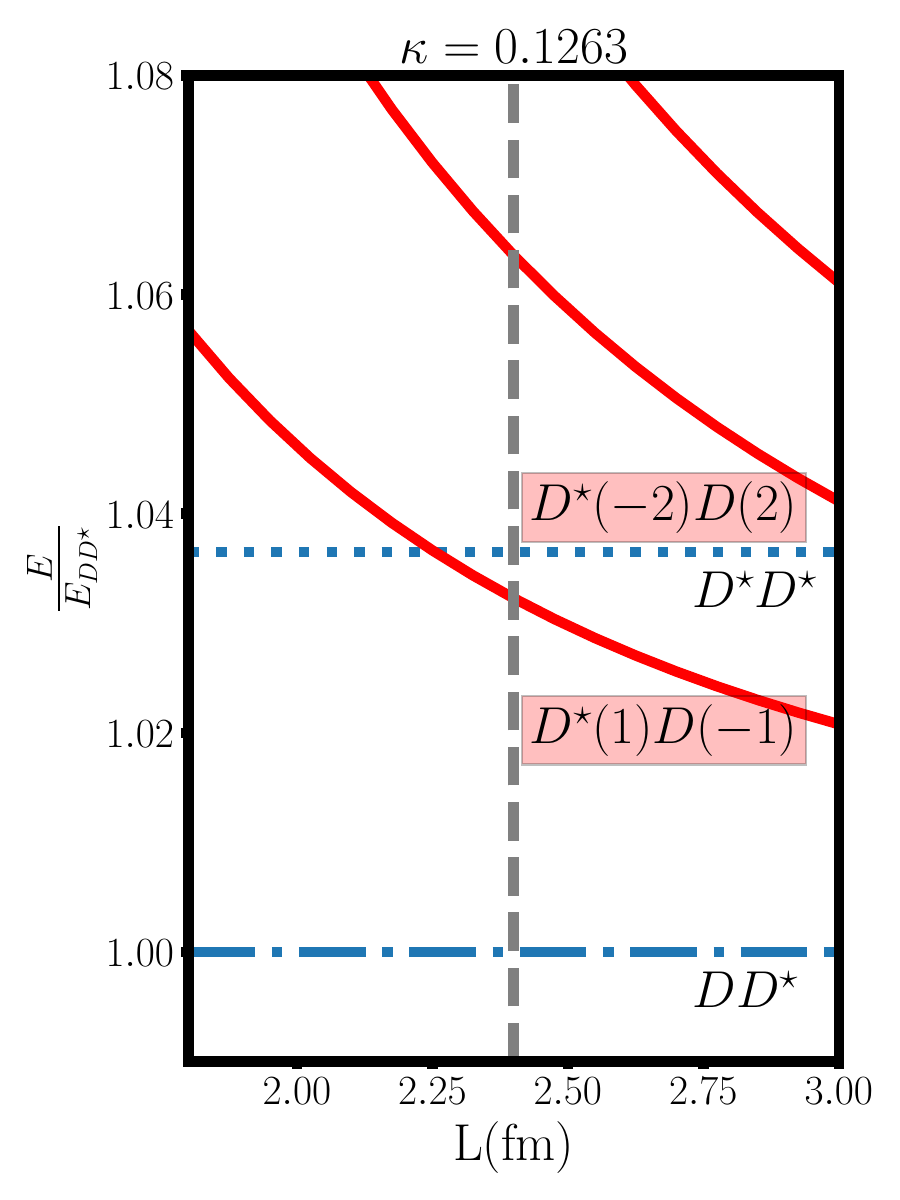}
\caption{$\kappa$=0.1263}
    \end{subfigure}
    \begin{subfigure}[t]{.32\textwidth}
\includegraphics[scale=0.30]{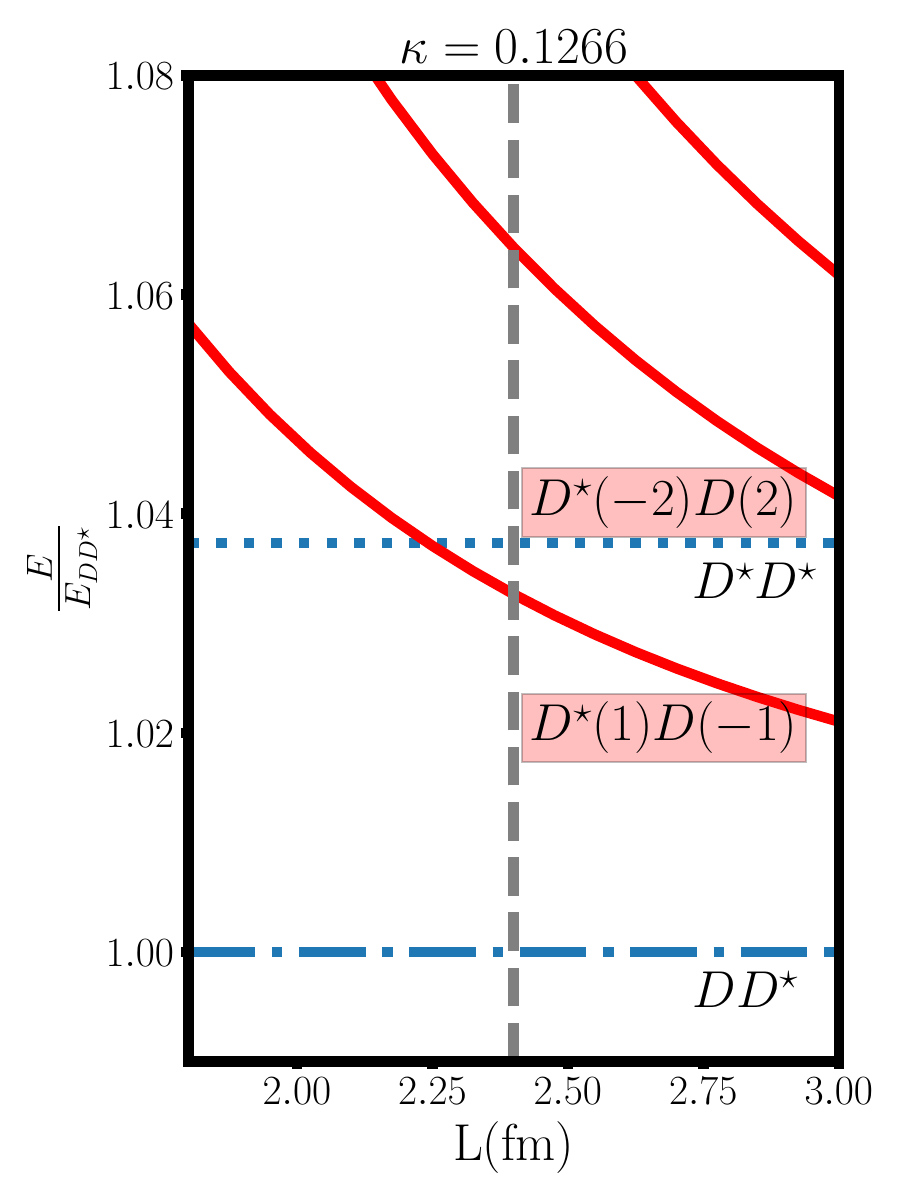}
\caption{$\kappa$=0.1266}
    \end{subfigure}%
    \begin{subfigure}[t]{.32\textwidth}
\includegraphics[scale=0.30]{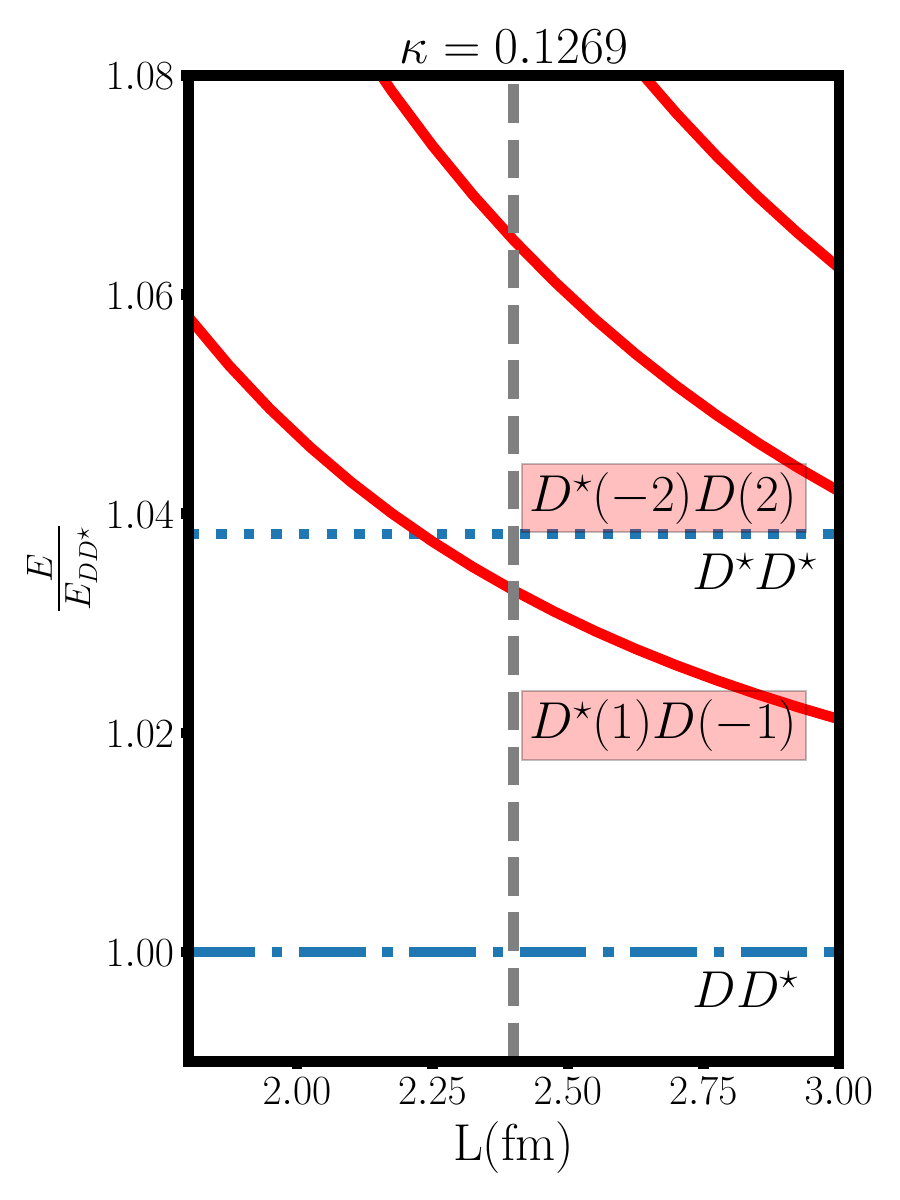}
\caption{$\kappa$=0.1269}
    \end{subfigure}%
    \begin{subfigure}[t]{.3\textwidth}
\includegraphics[scale=0.30]{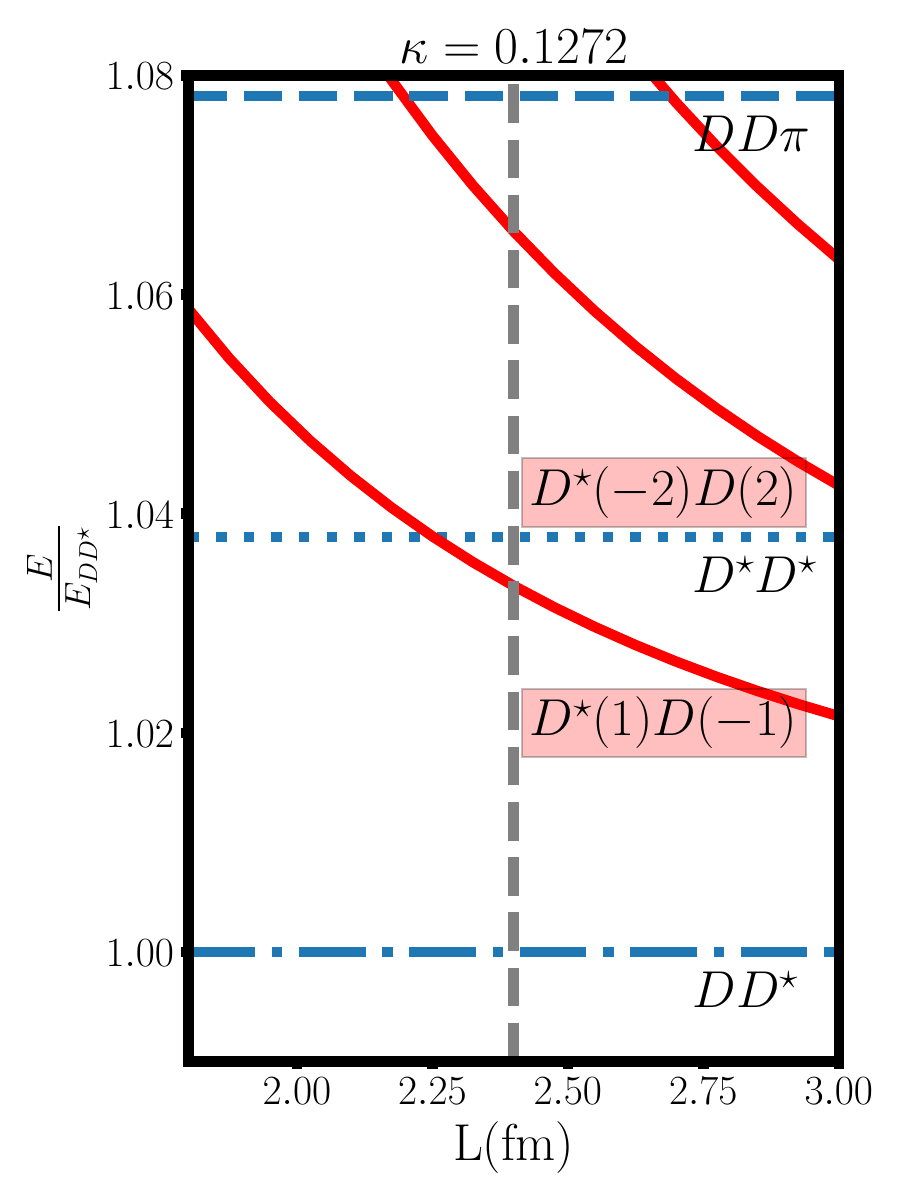}
\caption{$\kappa$=0.1272}
    \end{subfigure}
    \begin{subfigure}[t]{.3\textwidth}
\includegraphics[scale=0.30]{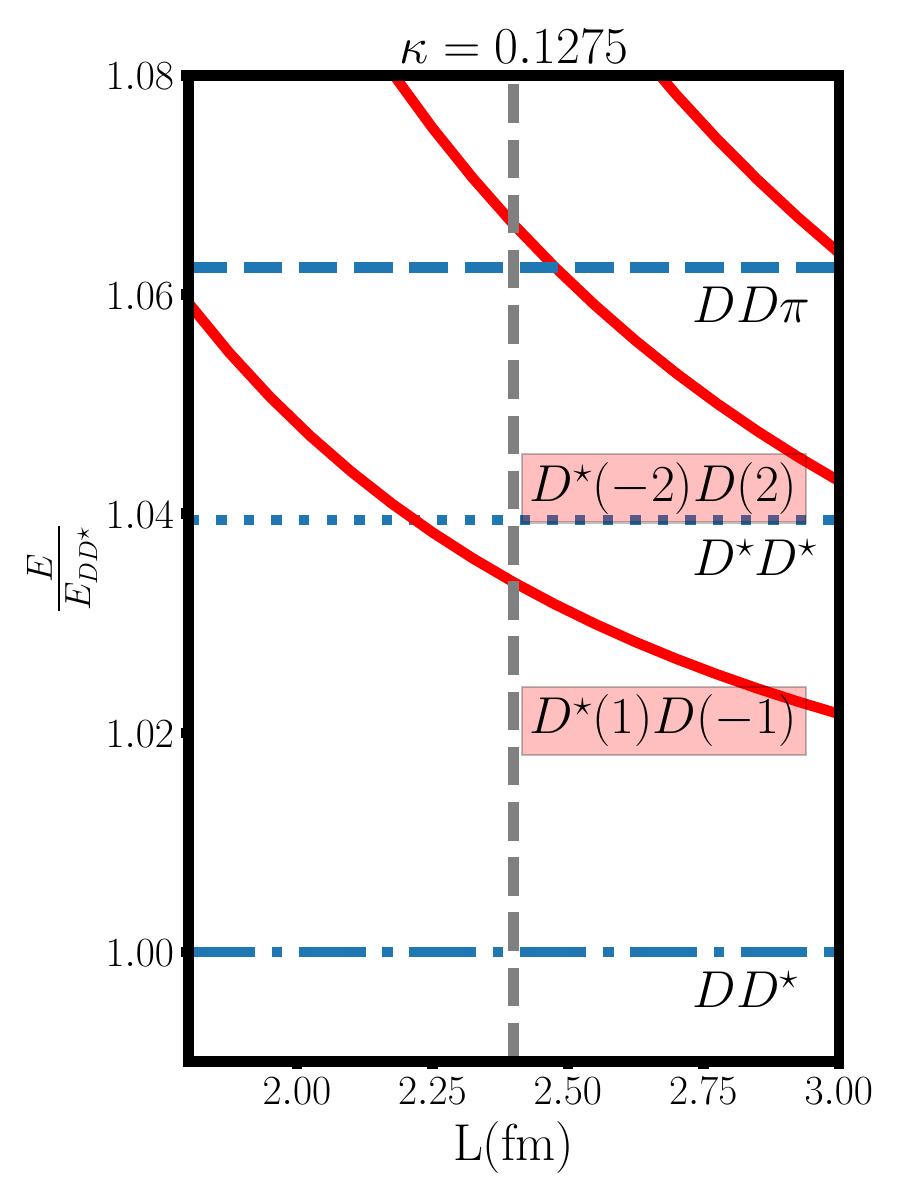}
\caption{$\kappa$=0.1275}
    \end{subfigure}
\caption{Variation of thresholds with changing light quark $\kappa$
from $m_\pi=688.8 (= m_{\eta_s})$ to $m_\pi = 400$ MeV.}
\label{fig:threshold1}
\end{figure}

In figure \ref{fig:threshold1} we show the trajectory of thresholds
for different $\kappa$'s  starting from the onset of elastic $D^\ast
D$  threshold up until the first inelastic $D^\ast D^\ast$ threshold.
This is the relevant domain of the L\"uscher formalism where it is
valid. We also depict the back-to-back non-interacting
$D^\ast  D$
energy level which is in the relevant domain. The vertical line
denotes the $16^3\times48$ lattice volume we use in our calculation.
When we perform our GEVP analysis we expect to find the energy levels
close to the intersection points of the vertical line. The presence
of valid non-interacting level in the relevant domain confirms that
we have the necessary finite volume setup to reliably perform the
$T_{cc}$ analysis and extract its pole trajectory from L\"uscher
formalism.

\begin{figure}[!ht]
\centerline{\includegraphics[width=0.85\textwidth,
height=0.35\textheight]{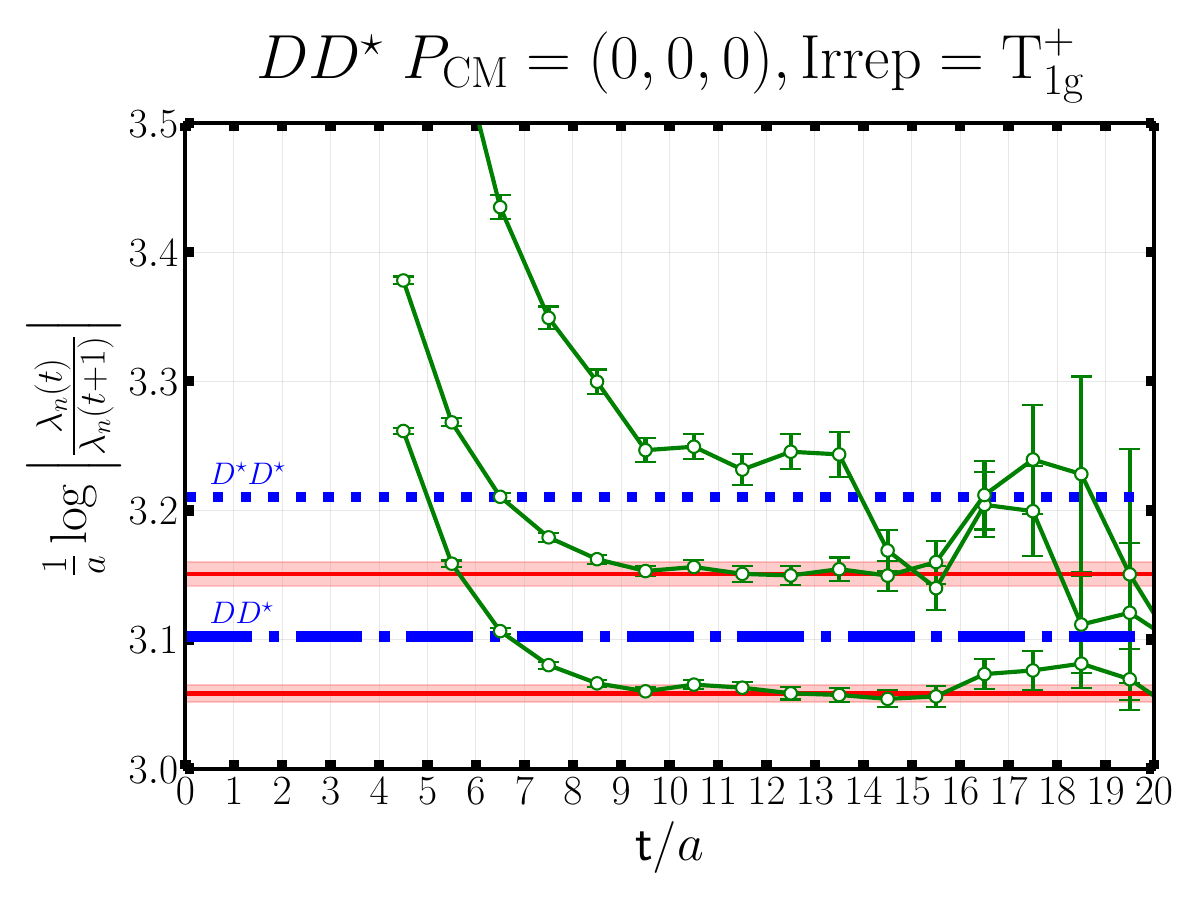}}
\caption{Energy levels of the GEVP matrix for light quark
$\kappa=0.12566$.}
\label{fig:gevp_plot}
\end{figure}
In Fig.~\ref{fig:gevp_plot} we present our preliminary GEVP
results
on the $16^3\times 48$ ensemble with light quark
$\kappa=0.12566$.
The correlator matrix has been constructed after projecting the
correlators into the $T_1^+$ irrep. We use time reversal symmetry
to average the forward and backward running principal correlators.
The principal correlators are fitted with single exponential
ans\"atz to obtain the finite volume spectrum. We plot the $3$
lowest lying levels below the inelastic $D^\ast D$ threshold.
The fit results are consistent with Fig. 3 in
Ref.~\cite{Prelovsek:2025vbr}. Following the
Ref.~\cite{Prelovsek:2025vbr}, we can discard the third energy
level close to the $D^\ast D^\ast$ threshold. The energy shift
from the non-interacting energy levels are a clear indication
of the presence of a interaction between the $D$ and $D^\ast$.

\section{ Summary and Outlook}
In this work we study the spectrum of doubly charm tetraquark $T_{cc}$
using MILC $N_f=2+1+1$ HISQ configurations. We include three set of
operators, namely diquark-antidiquark, molecular and scattering and
perform GEVP analysis on $5 \times 5$ correlation matrix. For actions,
we use Wilson-Clover for up/down and strange and anisotropic clover RHQ
for charm. We find that for $\kappa = 0.12566\; (m_\pi = 688.8 \; \text{MeV})$
the ground state of $T_{cc}$ lies below the $D^\ast D$ threshold, the
first and the second excited states appear to lie below and
above the $D^\ast D^\ast$ thresholds. The ground state likely
indicates an attractive
potential. It would be interesting to see what happens at lower pion mass
$\kappa = 0.1275\; (m_\pi = 400.4\; \text{MeV})$ and at heavier quark
mass points.

\section*{Acknowledgement}
We acknowledge financial support from the Department of Atomic Energy
(DAE), Government of India, and IMSc, Chennai. S.P. was partially
supported by DOE Grant KA2401045. We thank the MILC Collaboration
for providing the HISQ gauge ensembles used in this work.
Simulations have been carried out on the Bihan cluster of School
of Physical Sciences, NISER and the Kamet cluster at IMSc. P.M.
gratefully acknowledges support from the Department of Science and
Technology, India, SERB Start-up Research Grant No. SRG/2023/001235.
We also thank Prof. M. Padmanath of IMSc for fruitful discussions.

\end{document}